\documentclass[showpacs, showkeys, superscriptaddress]{revtex4}
\usepackage{graphicx}
\usepackage{hyperref}

\begin{document}
\title{Periodic orbit analysis for the deterministic path-preference traffic flow cellular automaton}
\author{Yoichi Nakata}
\email[Corresponding author; ]{ynakata@ms.u-tokyo.ac.jp}
\affiliation{Isotope Science Center, the University of Tokyo, 2-11-16 Yayoi, Bunkyo-ku, Tokyo 113-0032, Japan}
\author{Yoshihiro Ohta}
\affiliation{Institute of Biology and Mathematics (iBMath), Interdisciplinary Center of Mathematical Sciences, Graduate School of Mathematical Sciences, the University of Tokyo, 3-8-1 Komaba, Meguro-ku, Tokyo 153-8914, Japan}
\author{Sigeo Ihara}
\affiliation{Institute of Biology and Mathematics (iBMath), Interdisciplinary Center of Mathematical Sciences, Graduate School of Mathematical Sciences, the University of Tokyo, 3-8-1 Komaba, Meguro-ku, Tokyo 153-8914, Japan}
\affiliation{Research Center for Advanced Science and Technology, the University of Tokyo, 4-6-1 Komaba, Meguro-ku, Tokyo 153-8904, Japan}
\begin{abstract}
The path-preference traffic flow cellular automaton is suggested to model the dynamics of transcription. The main difference from the simple traffic flow model is that it contains another preferential paths at some sites. In this paper, we propose an exact analysis for the simplest version of this model.  We find that the number of particles is dominant to the dynamics of this cellular automaton and observed that there are not only expected phase shift but also several gaps as the number of particles increases. By considering the behavior of periodic orbits, we also determine the point where such gaps in the flow appear and the exact value of the flow.
\end{abstract}
\maketitle

\section{Introduction}
Discrete dynamical systems with discrete dependent variables are called cellular automata. They contain very rich structures and are applied to various area even if they have simple time evolution rules. Elementary Cellular Automaton (ECA) is one of good examples. This system consists on a sequence of sites and each site can take only two values and the time evolution of a site is defined only by the nearest neighbor sites \cite{Wolfram}. However, the ECA forms interesting time-space patterns and similar dynamics appears in various fields (for instance \cite{BelitskyFerrari}). 

The asymmetric simple exclusive process (ASEP) is a stochastic cellular automaton on one dimensional and originally presented as a model representing the dynamics of ribosomes in translation on RNA \cite{MacDonaldGibbsPipkin}, which is currently famous as the cellular automaton model representing the dynamics of traffic flow \cite{Derrida}. The model restricting the direction of the movement in the ASEP is called the TASEP and the deterministic version of the TASEP is the ECA rule 184, which is directly related to the differential equation expressing car traffics through the limiting procedure called ``ultradiscretization" \cite{TokihiroTakahashiMatsukidairaSatsuma}.

It is known that several phases exist in the states of the TASEP after a sufficient time has passed and such phases depend on the parameters of the system. In that of the ECA 184, there are two phases where particles move freely or make a traffic jam and it is known that the density of particles determines the phases \cite{SchadschneiderSchreckenberg}.

The path-preference cellular automaton model proposed by one of the authors and Nishiyama et al. should be interpreted as a model introducing a shortcut (or branching and jointing) for the TASEP and to explain the behavior of RNA Polymerase II in the  transcription dynamics of eucaryotes \cite{OhtaNishiyama}. Shortcuts mean that RNAPII will skip over the tracks that should be inherently due to the spatial proximity of the DNA placed in the high dimensional space and diffusion effect. Regulating the expression of function by the spatial structure of DNA has been actively studied particularly in recent years\cite{Cook, PapantonisLarkinWadaOhtaIharaKodamaCook, FraserBickmore}. On the other hand, to express transcription dynamics, several cellular automata model of RNAPII dynamics have been presented \cite{OhtaKodamaIhara, TripathiChowdhury, TripathiSchutzChowdhury}. Therefore, this model is a cellular automaton model to describe the dynamics of RNAPII in the three dimensional space. In this paper, we deal with a deterministic version of this model (i.e. an extension of ECA 184).

The existence of such shortcuts make the behavior of the system even more complicated and interesting. Numerical simulation confirms that discontinuous change in flow rate occurs by changing the number of particles, which is a major difference from the behavior of the ECA 184 \cite{BrankovPeshevaBunzarova, Negami}. Our target is to reveal the reason of such a phenomenon.

In this paper, we focus on periodic orbits of this system. For such periodic orbits, the flow rate can be computed exactly and we confirmed that it agrees with the numerical simulation. Furthermore, we find that the discontinuous change in the flow rate is due to transition to another class of periodic orbits by breaking the condition where the periodic orbit should hold.

Different from the ECA 184, orbits are found to be possibly attracted by different types of periodic orbits for some configurations even when the number of particles is the same. Although these orbits are unstable by the perturbations such as slightly altering the configuration of the particles, we have constructed a method to construct the trajectory. Furthermore, we have calculated the values such as the flow rate and the period of the orbit and discussed the conditions where such periodic orbits hold.

Conventionally, analysis using a statistical physics model has been actively performed for such a model \cite{TannaiNishinari}. We however propose the analysis by focusing on the orbit itself. In the analysis, we focus on the specific pattern through which orbits must pass, which is considered as an analogue of the Poincar\'e section in the dynamical system theory.

\section{Path-preference cellular automaton model}

\subsection{Modeling}

We first define the model which we deal with. This model is a special case of the path-preference cellular automaton model \cite{OhtaNishiyama}. 

The model is an extension of the ECA Rule 184. That is, basically the system consists of a finite amount of the sequence of boxes and a finite amount of balls and a box can contain at most one ball. A ball moves to the next box if there are no other balls. One time step evolution of this system is done when all ball move or stay. The destination of the ball at the last site is the first one, which means the periodic boundary condition. We denote the number of boxes and balls as $N$ and $M$, respectively and assign the number for each box in a sequential order.
However, at some sites the balls can also jump to the remote sites, not only the next sites. We denote $\epsilon_k$ as the site number of such a site and $\iota_k$ as that of jump destination sites from $\epsilon_k$-th site. We also assume that $\epsilon_k$ and $\iota_k$ have the relation:
\[
	1 \le \iota_K < \epsilon_1 << \iota_1 < \epsilon_2 << \iota_2 < \ldots < \epsilon_K << N,
\]
where $x << y$ means $y - x \ge 2$. We can set $\iota_K = 1$ without loss of generality because of the periodic boundary condition.

By setting these bifurcations, there are possibilities such that a ball can go to several destination and several balls are going to the same site at some time. For such cases, we introduce the following rules:
\begin{itemize}
\item The ball at $\epsilon_k$-th box jumps to $\iota_k$-th one when there are no balls at $\iota_k$-th (even if there is a ball at $\iota_k-1$-th).
\item The ball at $(\iota_k - 1)$-th box moves to $\iota_k$-th one only when there are no balls at $\iota_k$-th and $\epsilon_k$-th. 
\end{itemize}

\begin{figure}
\begin{center}
\includegraphics{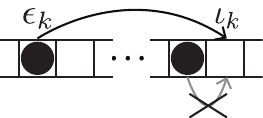}
\end{center}
\caption{The ball at $\epsilon_k$-th site jumps to $\iota_k$-th site if there are no balls and the ball at $(\iota_k-1)$-th box cannot move to the next if a ball jumps to there.}
\begin{center}
\includegraphics{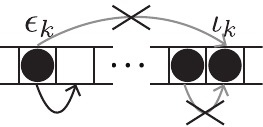}
\end{center}
\caption{The ball at $\epsilon_k$-th site jumps to the next if there is a ball at $\iota_k$-th and there are no balls at the next site.} \label{fig-ex2}
\end{figure}

This rule indicates that we set a priority for balls jumping via shortcut. We call this rule ``jump priority". Inversely, we can also set the priority for balls moving to next sites. We call this rule ``move priority". In this paper, we employ jump priority just because of the mathematical symmetry described as follows but not to reflect the real dynamics which the original model tends to represent.

With the definitions above, let us introduce the mathematical expression of the time evolution rule of this system. We set $u^t_j$ the number of balls (which takes $0$ or $1$) at the $j$-th site in time $t$ and $f(k, l)$ the number of transiting balls from $k$-th site to $l$-th. Hereafter, we use the letter $j$ for indicating the site number. Due to the time evolution rule, the movement of a ball to the next site is expressed as
\begin{equation}
	f(j, j+1) = \min(u^t_j, 1-u^t_{j+1}) \label{defmoveparticle}
\end{equation}
for $j \neq \epsilon_k$ ($k=1, 2, \ldots, K$). 

Since we introduce the jump priority rule, the number of jumping balls from $\epsilon_k$-th to $\iota_k$-th is
\begin{equation}
	f(\epsilon_k, \iota_k) = \min(u^t_{\epsilon_k}, 1-u^t_{\iota_k}).
\end{equation}
In the case $f(\epsilon_k, \iota_k) = 1$, those of moving balls are written in
\begin{eqnarray}
	f(\epsilon_k, \epsilon_k+1) = 0 \\ 
	f(\iota_k-1, \iota_k) = 0
\end{eqnarray}
and in the case $f(\epsilon_k, \iota_k) = 0$, 
\begin{eqnarray}
	f(\epsilon_k, \epsilon_k) = \min(u^t_{\epsilon_k}, 1-u^t_{\epsilon_k+1}) \\ 
	f(\iota_k-1, \iota_k) = \min(u^t_{\iota_k-1}, 1-u^t_{\iota_k}).
\end{eqnarray}
Here, we consider the modulo of $N$ for the index of the site number.

Then, we obtain the time evolution rule:
\begin{eqnarray}
	u^{t+1}_{\epsilon_k} = u^t_{\epsilon_k} + f(\epsilon_k-1, \epsilon_k) - f(\epsilon_k, \iota_k + 1) - f(\epsilon_k, \epsilon_k+1) \\
	u^{t+1}_{\iota_k} = u^t_{\iota_k} + f(\iota_k-1, \iota_k) + f(\epsilon_k, \iota_k) - f(\iota_k, \iota_k+1)
\end{eqnarray}
for $k=1, \ldots, K$ and
\begin{equation}
	u^{t+1}_j = u^t_j + f(j-1, j) - f(j, j+1).
\end{equation}
for other sites.

By adding this formula for $j=1, \ldots, N$, one has $\sum_{j=1}^N u^{t+1}_j = \sum_{j=1}^N u^t_j$, which means the conservation law of the total number of balls. Furthermore, when introducing a new dependent variable $v^t_j := 1 - u^t_j$, the time evolution rule of $v^t_j$ is the same as that of $u^t_j$ under the replacement $j \to -j$, which means that the dynamics is the same under the inverted direction when interchanging empty sites and occupied ones. Such a good symmetry does not hold if one sets the move priority rule. 

Since this model is deterministic and the states are finite amount, an orbit which starts from an arbitrary initial state is finally attracted by a periodic orbit (the steady states are regarded as the periodic orbits with period 1), which should be called as the limit cycle.

\section{Analysis of the simplest case: $K=1$}

We first consider the simplest case $K=1$. In this case, the site where a ball can shortcut is only $j = \epsilon_1$ and the destination site is $j = \iota_1 = 1$. We call the sites from $j=1$ to $j = \epsilon_1$ the main region and those from $j=\epsilon_1 + 1$ to $j = N$ the skipped region. We also define $N_m$ and $N_s $ as the length of the main region and that of the skipped region, respectively.

\subsection{Aspect of the system}

\begin{figure}
\includegraphics[scale=0.5]{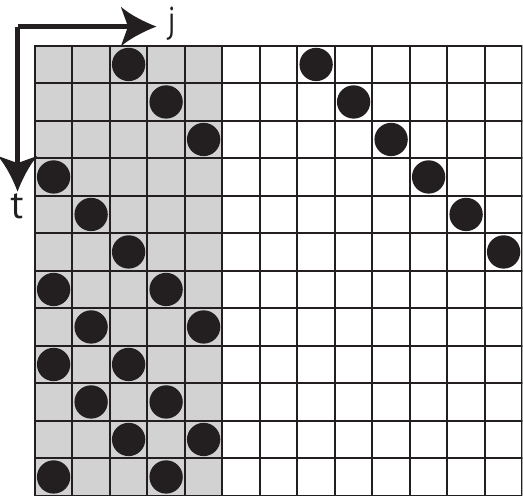}
\caption{An example of general orbits for $N_m = 5$, $N_s = 8$, $M = 2$. The sites of the main region are gray-meshed.} \label{examplegeneral}
\end{figure}

We first consider the behavior of this system by focusing on the number of balls $M$. In the case that $M$ are sufficiently small, all balls are finally trapped by the main region once they arrived there for most of initial values (See Fig. \ref{examplegeneral}). Such behavior can be kept while $M \le N_m/2$. If $M > N_m/2$, the main region cannot hold all balls and overflowed ones are emitted to the skipped region. In other words, the traffic on the main region is always optimized for maximizing its flow efficiency. In particular, if $N_m$ is even, balls in the main region generate a closed flow which prohibits other balls to join and those in the skipped region never come back to the main region and finally cause the unsolvable traffic jam at the end of the skipped region. If $N_m$ is odd, there is at least one area which satisfies $u^t_j = 0$ and $u^t_{j+1} = 0$ in the main region for any limit cycles. At space-time pattern, this area looks like a notch, which permits the balls at the skipped region to move back to the main region and helps the dynamics be richer. In this paper, we mainly consider the case $N_m$ is odd.

\begin{figure}
\includegraphics[scale=0.5]{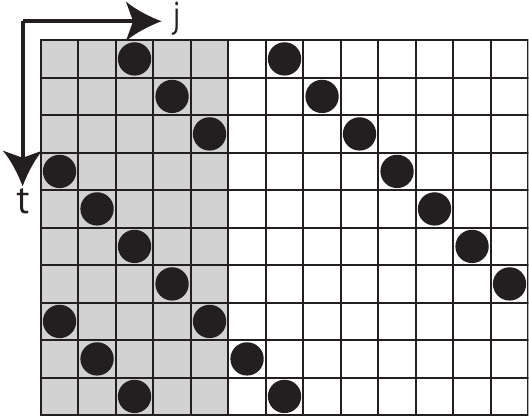}
\caption{An example of special periodic orbits. The system parameters are the same as those in Fig. \ref{examplegeneral}.} \label{examplespecial}
\end{figure}

The limit cycles starting from most of the initial states have certain behaviors determined by the number of balls $M$. However, some limit cycles starting from some special initial states show behaviors out from them. 
For example, in the time space pattern in Fig. \ref{examplespecial}, one ball is not trapped by the main region and goes around the skipped region even if the system satisfies $M \le N_m/2$. Such orbits occur because the configurations are well balanced and are easily attracted by other orbits with a little perturbation such as shifting one of balls to the neighbor empty site. We call these unstable orbits ``special periodic orbits" and we call the orbits which are started from most of initial states and robust for little perturbation as ``general periodic orbits".

\subsection{Analysis focusing on flow}

\begin{figure}
\begin{center}
\includegraphics[scale=0.5, angle=90]{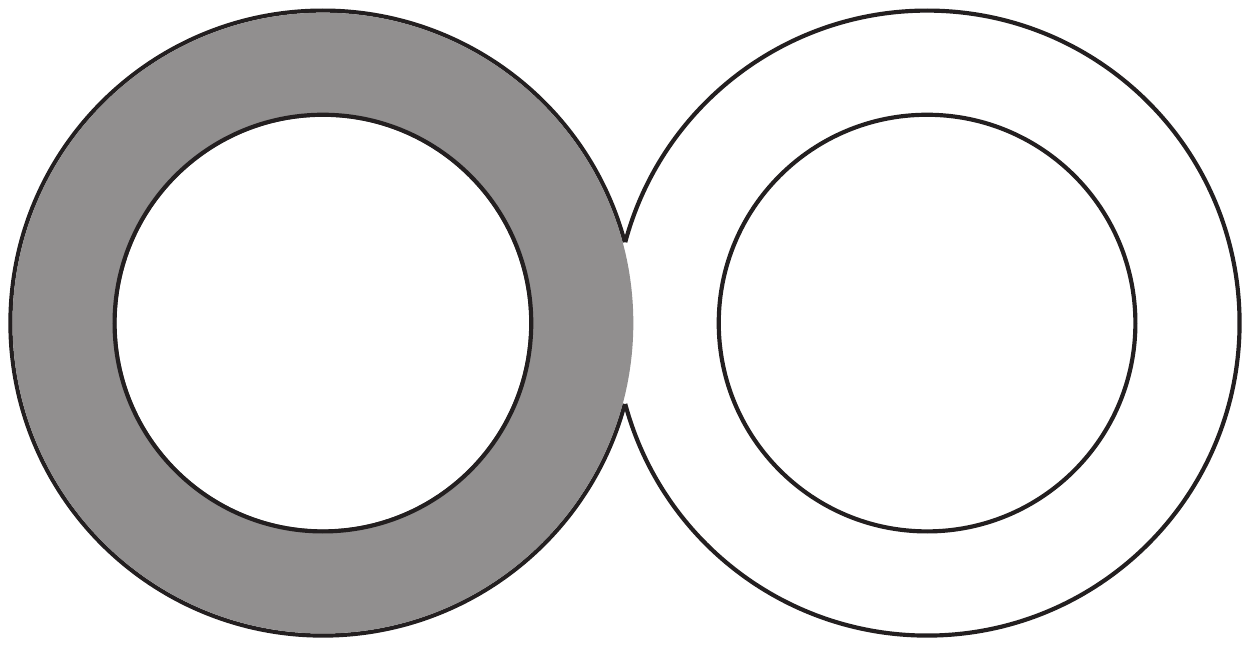}
\caption{An expression of one main and skipped regions model.} \label{eightshape}
\end{center}
\end{figure}

The value ``flow", which is defined as the number of moving balls per a site, is a powerful tool for analyzing the behavior of the general traffic models. However, it is not naively applicable for this model because there is an ambiguity of the distance for the jump. If one wants to match a model with experiments, we should take this distance properly. For example, it should be $1$ for the figure of 8-shape form like Fig. \ref{eightshape}  and it is taken as the length of the hurdled intron in the original path-preference model. In this paper, we consider the flow of each region instead of the flow of whole the system to focus on mathematical properties. We also ignore the movement between two regions. Then, we define $J_m$ and $J_s$ as the flow of the main region and the skipped one respectively, which are written in
\begin{eqnarray}
	J_m = J_m(t) := \frac{1}{N_m-1} \sum_{j=1}^{N_m-1} f(j, j+1) \\
	J_s = J_s(t) := \frac{1}{N_s-1} \sum_{j=N_m+1}^{N-1} f(j, j+1).
\end{eqnarray}
We also define the time average of value $X$($= J_m, J_s$) by
\begin{eqnarray}
	\langle X \rangle = \lim_{T \to \infty} \frac{1}{T} \sum_{t=0}^{T-1} X(t). 
\end{eqnarray}
Now, remember that arbitrary orbit is attracted by a periodic orbit. By defining the period of the orbit as $T$, one can rewrite
\begin{eqnarray}
	\langle X \rangle = \frac{1}{T} \sum_{t=t_0}^{t_0 + T - 1} X(t) \label{deftimeaverage}
\end{eqnarray}
for sufficiently large $t_0$. We note that the period depends of the initial states, i.e. $T = T(\{ u^0_j \}_{j=1}^N )$.

\begin{figure}
\includegraphics[scale=0.8]{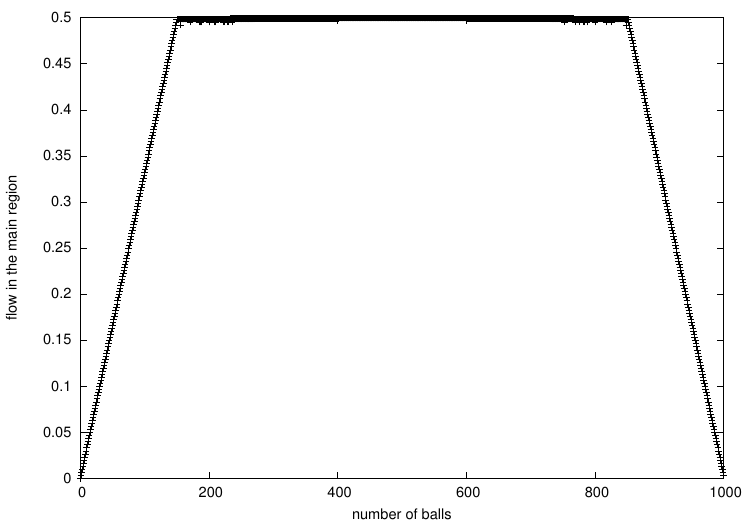}
\caption{The graph of the flow in the main region $\langle J_m \rangle$ for several initial state for each $M$ where $N_m = 301$ and $N_s = 700$.} \label{fig_flow_main}
\includegraphics[scale=0.8]{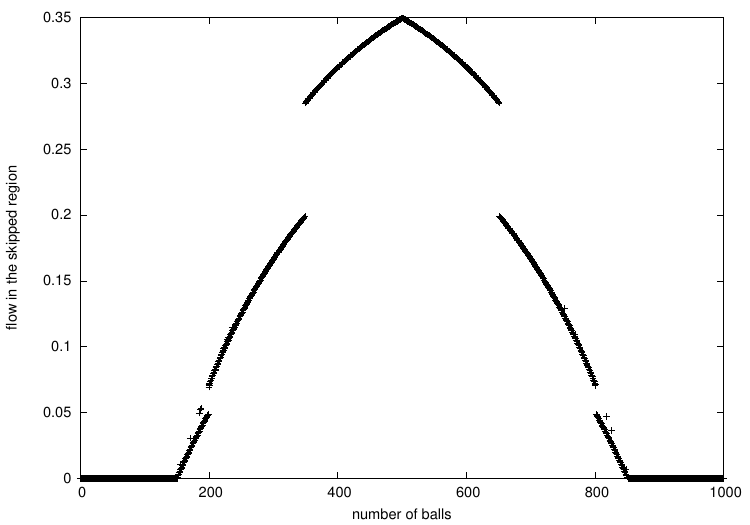}
\caption{The graph of the flow in the skipped region $\langle J_s \rangle$ for several initial state for each $M$ where $N_m = 301$ and $N_s = 700$.} \label{fig_flow_skipped}
\end{figure}

FIG. \ref{fig_flow_main} and \ref{fig_flow_skipped} are graphs relating between the flow $\langle J_m \rangle$, $\langle J_s \rangle$ and the number of balls $M$ for fixed $N_m$ and $N_s$ (we consider several initial states for each $M$). One can find that not only phase shift around $M = N_m/2$ discussed above but also several non-continuous gaps appear in the flow in the skipped region. One can also observe that the flow in the skipped region can take several value for some $M$. We note that the graph is symmetric with $M = N/2$ because of the symmetry due to the jump priority time evolution rule. To discuss the reason of such phenomena including the exact point of gaps and the flow of each part, we focus on several representative periodic orbits.

\bigskip

\subsection{Periodic orbit: class 1-0}

We first consider the case where $N$ is even and $M = N/2$ (i.e. $N_s$ is odd because we assume that $N_m$ is odd). In this case, arbitrary orbits are finally attracted by the limit cycle $u^t_j = t+j$ ($\mathrm{mod}\ 2$) with period $2$ and its time-space pattern forms a checkerboard one. The exact flow is
\begin{equation}
	\langle J_m \rangle = \langle J_s \rangle = \frac{1}{2} \label{flow1-0}.
\end{equation}

\subsection{Periodic orbit: class 1-1}

We next consider the case $M < N/2$, where the several sequential empty boxes appear in the limit cycles. We first focus attention on specific patterns in the skipped region of which one ball and one empty box appear alternately. We call such patterns ``clusters". Observation shows that there is a pair of sequential empty sites in the main region in most of time steps. We call such sites ``notch". The notch goes around in the main region.

The dynamics of periodic orbits are controlled by the number of clusters and the dynamics of the notch.
We first consider the case that the limit cycle contains only one notch and one cluster.

We pay attention to the time when the notch is the top of the main region. We assume that the initial state is  $u^t_j = 1$ ($j = 3, 5, \ldots, 2M-3, 2 M - 1$) and $u^t_j = 0$ otherwise. The time when the head of the cluster arrives at the end is $t_0 := N - 2 M -1$.
While balls and empty boxes appears alternatively at the last site of the main region in time evolution, the balls in the skipped region can go back to the main region. Then, when the top of the cluster in the skipped region, it makes traffic jam until the notch arrives the last site of the main region. We define this time as $t_1:=N_e-1$. In this case, one has the relation: $t_0 \le t_1$.

When the notch arrives there, the balls in the cluster go back to the main region. Next, when this ball arrives at the top of the main region, a ball is at the end of the main region and it cannot go to the top. Then, it is bumped out to the skipped region. In the next time, neither $j=0$ nor $j=N_e$ is occupied. Then, a ball of the cluster goes back again. These processes repeated until all of the balls of the cluster go back to the main region. After all balls go back, there appears a new notch in the top and a ball in the end can go around in the main region again.

The number of emitted balls from the main region is the same as the number of balls existing in the skipped region for the initial state. Then, the state of the system when a new notch are created is nothing but the initial state.
The time it takes for the balls making a jam to start going back to the main region and finish it is $2\{M-(N_e-1)/2\}$. Therefore, the period of this orbit is $T = t_1 + 2\{M-(N_e-1)/2\} + 1 = 2 M +1$.

Now, the number of balls emitted to the skipped region per a period $Q$ is the same as that in the skipped region at the initial state and written in 
\begin{equation}
	Q = M-\frac{N_e-1}{2}.
\end{equation}
The balls in the skipped region goes around this region just one time per a period. Then, the sum of the flow for whole the skipped region and a period is expressed as:
\begin{equation}
	\sum_{t=t_0}^{t_0 + T - 1} \sum_{j=N_m+1}^{N-1} f(j, j+1) = Q (N_s - 1).
\end{equation}
Then, the time average of the flow at the skipped region is expressed as
\begin{eqnarray}
	\langle J_s \rangle = \frac{1}{T} \sum_{t=t_0}^{t_0 + T - 1} \frac{1}{N_s-1} \sum_{j=N_m+1}^{N-1} f(j, j+1) = \frac{Q}{T} = \frac{2M-N_e+1}{2(2M+1)}
\end{eqnarray}
and the similar discussion yields the exact value of the flow at the main region by
\begin{equation}
	\langle J_m \rangle = \frac{2M}{2(2M+1)}.
\end{equation}
The condition $t_0 \ge t_1$ is written in
\begin{equation}
	\frac{N_i}{2} \ge M,
\end{equation}
which is the necessary condition where orbits can be attracted this limit cycle.

The flow of this mode corresponds the connected component of the graph which is the nearest to $M=N/2$.

\begin{figure}
\includegraphics[scale=0.5]{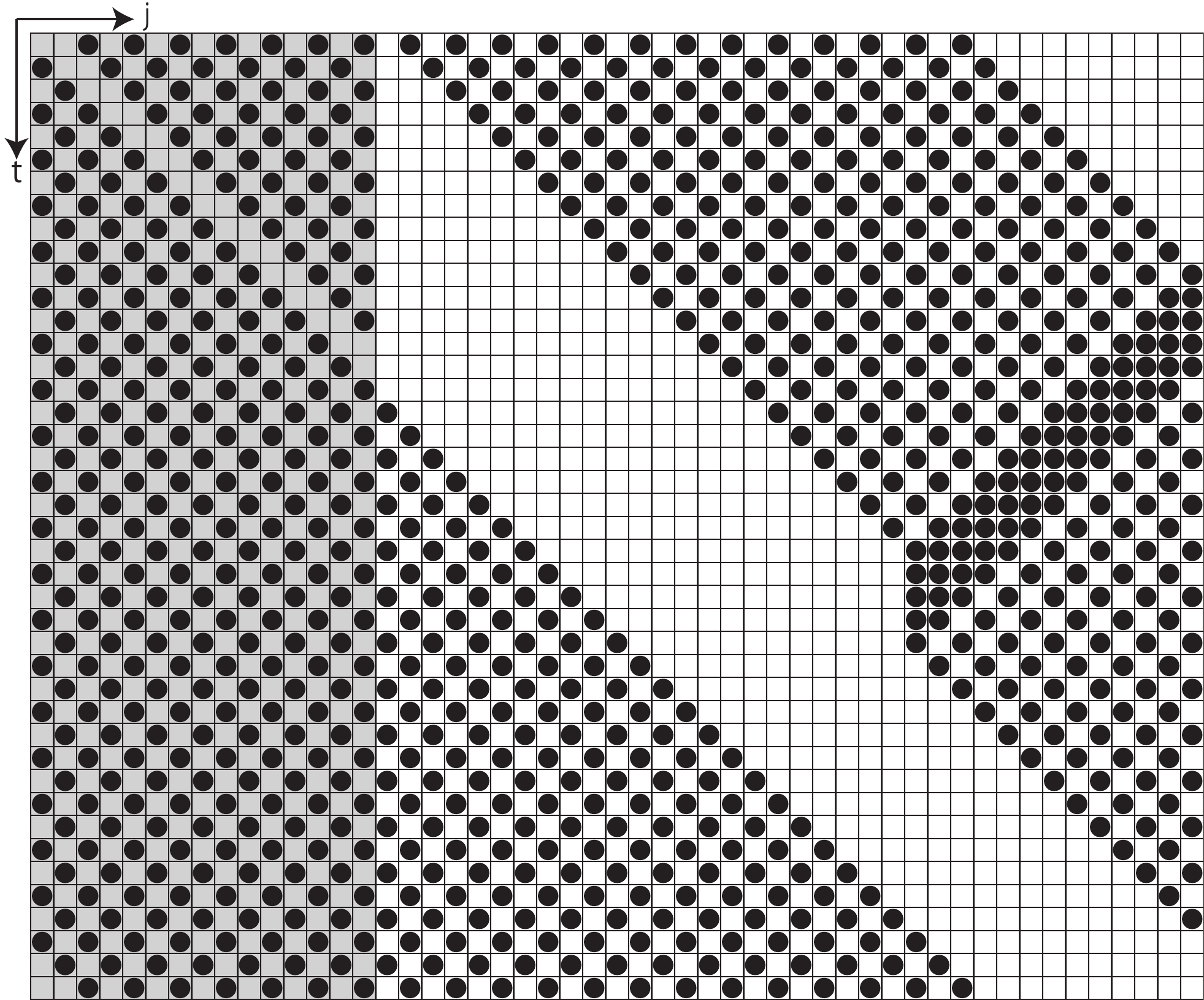}
\caption{An orbit of limit cycle class 1-1 for $1$ period. The parameters are $N_e = 15$, $N_i = 36$, $M = 20$.} \label{fig-smpl-time-evol-1}
\end{figure}

\subsection{Periodic orbit: class 1-2}

If $t_1 < t_0$, the head of the cluster at the the skipped region does not reach to the end at $t = t_0$. Then, the notch once goes back to the top of the main region and moves to the end again. The time when it arrives at the end again is at $t'_1 = 2 N_e + 1$. If $t'_1 \ge t_0$, the balls of the cluster move back to the main region at $t = t'_1 + 2$ and by the same discussion as above, the period is $T = 2 M + N_e + 1$ and the number of balls emitted to the skipped region per a period $Q = M-\frac{N_e-1}{2}$. Therefore, the flow of each region is written in
\begin{eqnarray}
	\langle J_s \rangle = \frac{Q}{T} = \frac{2M-N_e+1}{2(2M+N_e+1)} \label{flow-i-1-2} \\
	\langle J_m \rangle = \frac{2M+N_e-1}{2(2M+N_e+1)} \label{flow-e-1-2}
\end{eqnarray}
and the condition $t'_1 \ge t_0 > t_1$ is 
\begin{equation}
	\frac{N_i - N_e}{2} \le M < \frac{N_i}{2}. \label{cond-1-2}
\end{equation}
We next consider the initial state $u^t_j = 1$ ($j = 3, 5, \ldots, N_e, N_e+2, \ldots, 2 m_1 + N_e - 2, 2 m_1 + N_e, 2 m_1 + 2 N_e + 2, \ldots, 2 m_1 + 2 m_2 + 2 N_e$) and $u^t_j = 0$ for other sites, where $m_1$ and $m_2$ are positive number satisfying $m_1 + m_2 + (N_e-1)/2 = M$. In this initial state, there are two clusters with $m_1$ and $m_2$ balls respectively in the skipped region. The head of the top cluster arrives at $j = N$ in $t_0= N - 2 m_1 - 2 m_ 2 - 2 N_e - 1$ and the notch arrives at $t_1 = N_e - 1$. If $t_0 < t_1$, by the same discussion as above, the head block of the skipped region moves to the main region and $m_2$ balls of the main region are emitted to the skipped region. Then after passing $t = 2 m_2 + t_0 + 1$, the state is $u^t_j = 1$ ($j = 3, 5, \ldots, N_e, N_e+2, \ldots, 2 m_2 + N_e - 2, 2 m_2 + N_e, 2 m_2 + 2 N_e + 2, \ldots, 2 m_2 + 2 m_1 + 2 N_e$) and $u^t_j = 0$ for other sites, which is the same as the initial state by interchanging $m_1$ and $m_2$. Then, by repeating this procedure again, we can show that the orbit is periodic with period $T = 2M + N_e + 1$ and the flow is
\begin{eqnarray}
	\langle J_s \rangle = \frac{m_1 + m_2}{T} = \frac{2M-N_e+1}{2(2M+N_e+1)} \\
	\langle J_m \rangle = \frac{2M+N_e-1}{2(2M+N_e+1)}.
\end{eqnarray}
The condition where this periodic orbit holds is:
\begin{itemize}
	\item the time when the top cluster arrives at the end of the skipped region is earlier than that when the notch arrives at the end of the main region.
	\item the time when all of the balls of the top cluster have gone back to the main region is earlier than the time when the second top cluster arrives at the end of the skipped region.
\end{itemize}
The first condition is $N-(2 m_1 + 2m_2 + 2 N_e) \le N_e - 1$, which is equivalent to $(N_i-N_e)/2 \le M$ and the second condition is $N_e + 1 + 2 m_2 + 1 \le N-(2 m_1 + 2 N_e)$, which is equivalent to $M \le (N_i-1)/2$ by virtue of the discussion in the previous section. Then, we obtain the condition:
\begin{equation}
	\frac{N_i - N_e}{2} \le M < \frac{N_i}{2}.
\end{equation}
The flow rates and condition are the same as (\ref{flow-i-1-2}), (\ref{flow-e-1-2}) and (\ref{cond-1-2}). Then, we found that these two orbits belong to the same class. Indeed, the first orbit is considered as one of the second type of orbits by setting $m_2 = 0$. We note that the real period is the half of $T$ if $m_1 = m_2$. However, the flow does not change because we double-count both the balls and periods.

\begin{figure}
\includegraphics[scale=0.5]{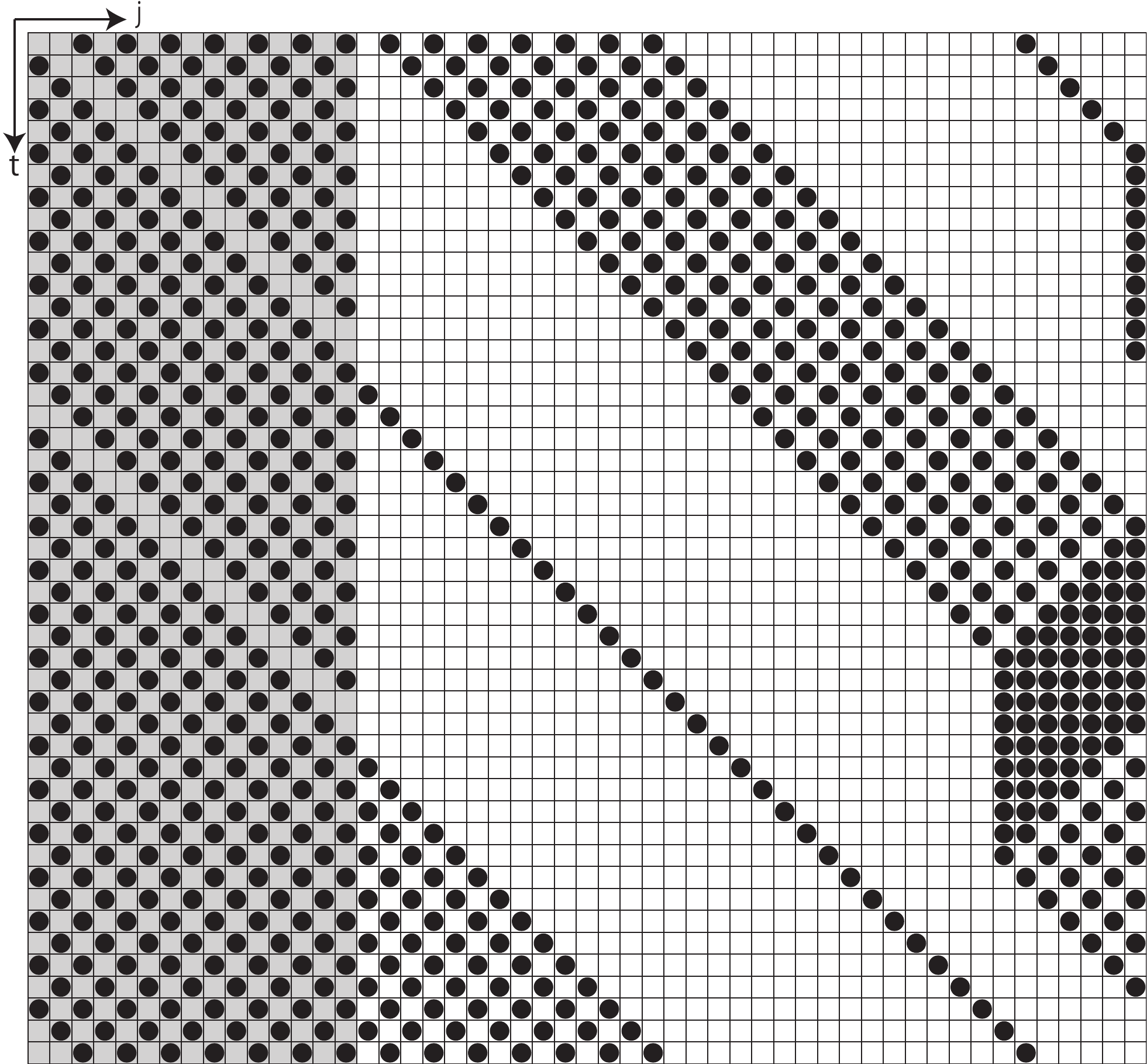}
\caption{An orbit of limit cycle class 1-2 for $1$ period. The parameters are $N_e = 15$, $N_i = 36$, $M = 15$.}
\end{figure}

\subsection{Periodic orbit: class 1-$\gamma$}

By generalizing the discussion in the previous section, we can show that the initial state $u^t_j = 1$ ($j=3, 5, \ldots, N_e, N_e+2 , \ldots, 2 m_1 + N_e, 2 m_1 + 2 N_e + 2, \ldots, 2 m_1 + 2 m_2 + 2 N_e, \ldots,  2 m_1 + \ldots + 2 m_{\gamma} + \gamma N_e + 2, \ldots, 2 m_1 + \ldots + 2 m_{\gamma} + \gamma N_e$) and $u^t_j = 0$ for other sites yields the periodic orbit whose properties are 
\begin{eqnarray}
	T = 2 M + (\gamma - 1) N_e + 1 \\
	\langle J_i \rangle = \frac{2M-N_e+1}{2(2M+(\gamma -1 ) N_e+1)} \label{flow-i-1-g} \\
	\langle J_e \rangle = \frac{2M+(\gamma-1)(N_e - 1)}{2(2M+(\gamma -1 ) N_e+1)}
\end{eqnarray}
for $\gamma > 0$ and the necessary condition to attract this limit cycle is
\begin{equation}
	\frac{N_i - (\gamma - 1) N_e}{2} \le M < \frac{N_i - (\gamma-2) N_e}{2}
\end{equation}
while $M > N_e/2$. Here, we stress that these conditions are disjoint and cover all $M \le N/2$, i.e., this class of limit cycles is uniquely determined by the number of balls if the initial state is general.

It should be noted that these formulae of flow depend only on the number of balls and clusters and the length of the main region but not the length of the skipped region and the condition depends only on them but not the detail of each cluster.

\begin{figure}
\includegraphics[]{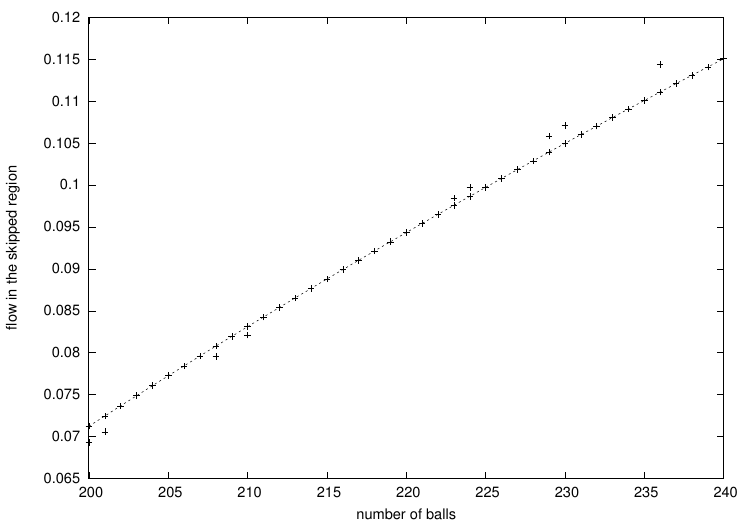}
\caption{A part of the graph of the flow in the skipped region for $N_m = 301$, $N_s = 700$. The dashed curve is depicted by the formula (\ref{flow-i-1-2}).  Points are numerically calculated by definition (\ref{deftimeaverage}).} \label{fig-expr-vs-theor-general}
\end{figure}

Fig. \ref{fig-expr-vs-theor-general} shows an example of the flow of each region versus the number of balls. We can observe that most of points are on the dashed curve which is depicted from the formula (\ref{flow-i-1-2}). This means that the orbits providing these values are general ones.

We mention that the total number of the balls which are bumped to the skipped region in a period is $Q = (2M-N_e+1)/2$, which is not depend on the mode and continuous by the number of balls. The value which changes discontinuously by the number of balls is period, which is depend on the mode. This is the reason why the flow changes discontinuously.

In the last of this section, we note that by setting $\gamma = 0$, the flow is equal to (\ref{flow1-0}). However, the period is $2M-N_e+1$, which is equal to twice of the number of balls in the skipped region. The real period $2$ is trivially an aliquot of this system period but this is a little interesting result.

\subsection{Special periodic orbits}

\begin{figure}
\includegraphics[scale=0.75]{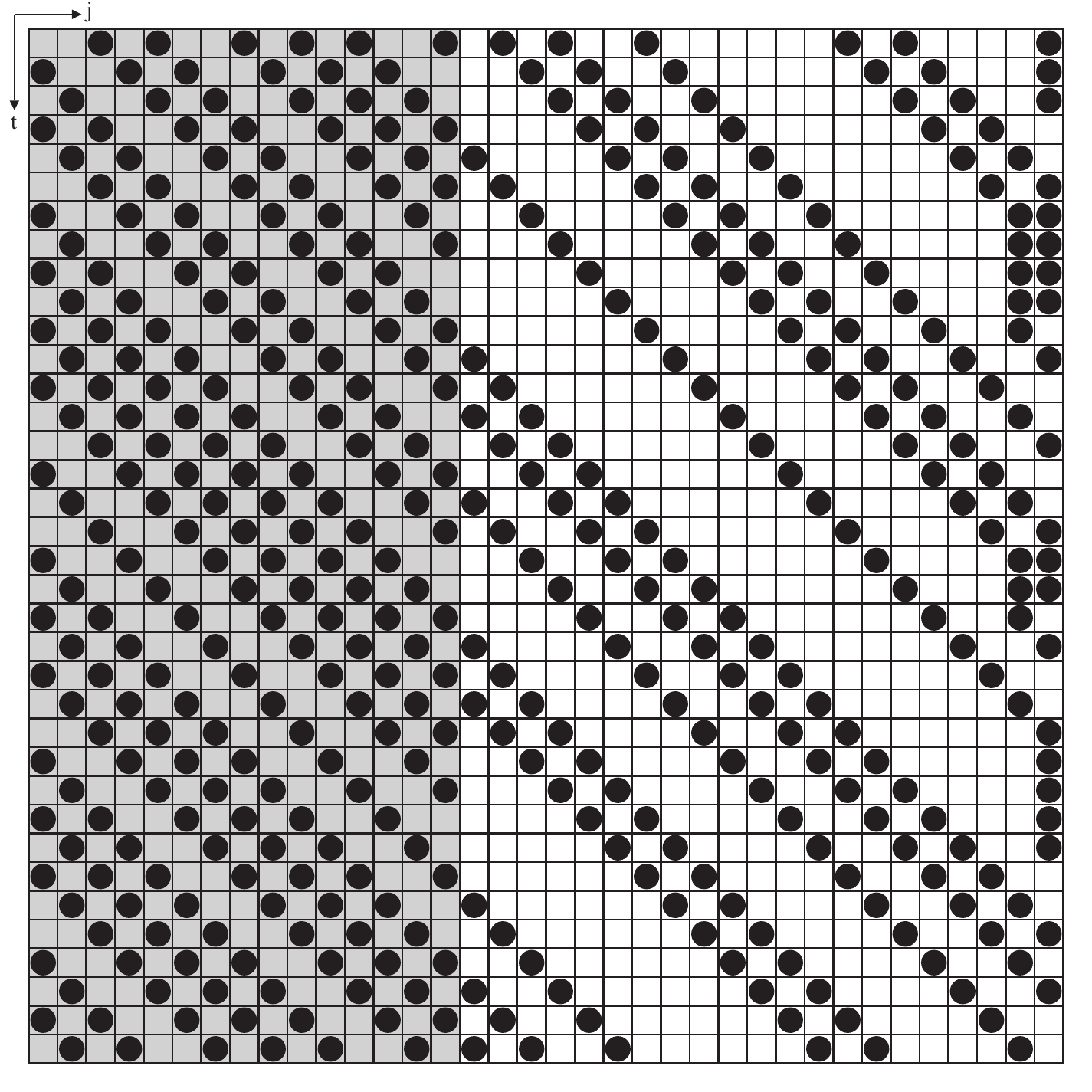}
\caption{An example of special periodic orbits. The parameters are $N_e = 15$, $N_i = 21$, $M = 12$. The period of this limit cycle is $36$ though that of general limit cycles should be $25$. We also note that the general limit cycles also exist for this parameter.} \label{fig-smpl-time-evol3}
\end{figure}

Next we consider special periodic orbits. The observation of such orbits (See Figure \ref{fig-smpl-time-evol3}) reveals that there are several notches in the main region, which is the major difference from the general periodic orbits. Obviously, the parity of the number of notches and $N_e$ are the same.

\begin{figure}
\begin{tabular}{c}
\includegraphics[scale=0.5]{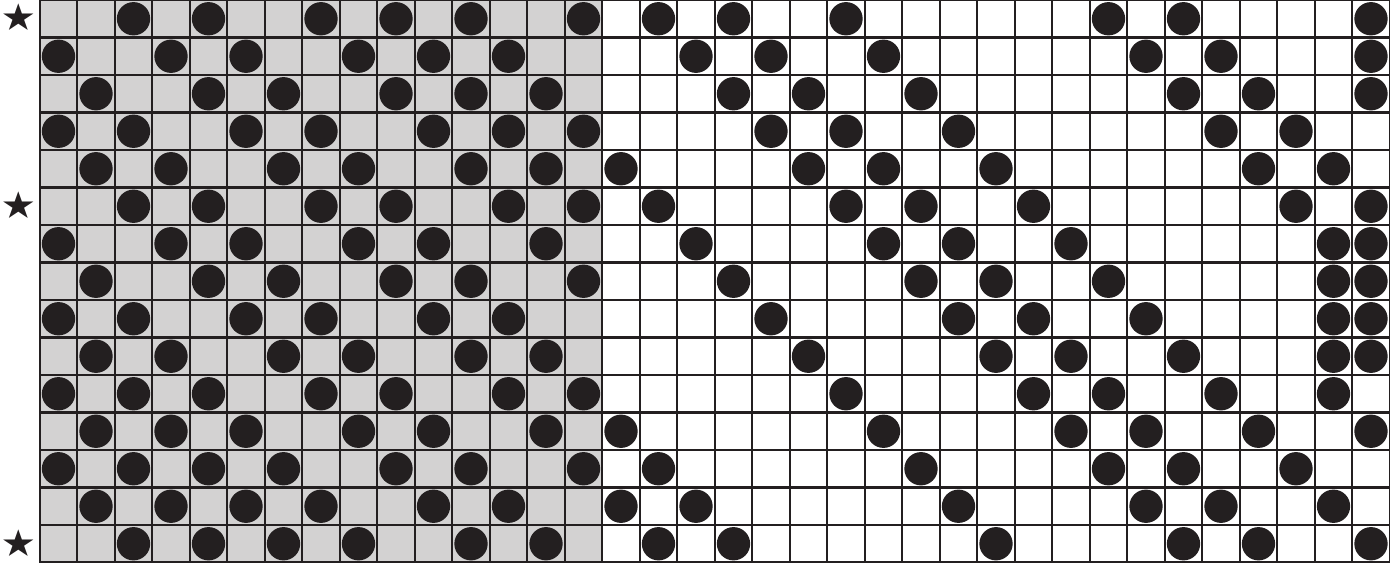} \\
\\
\includegraphics[scale=0.5]{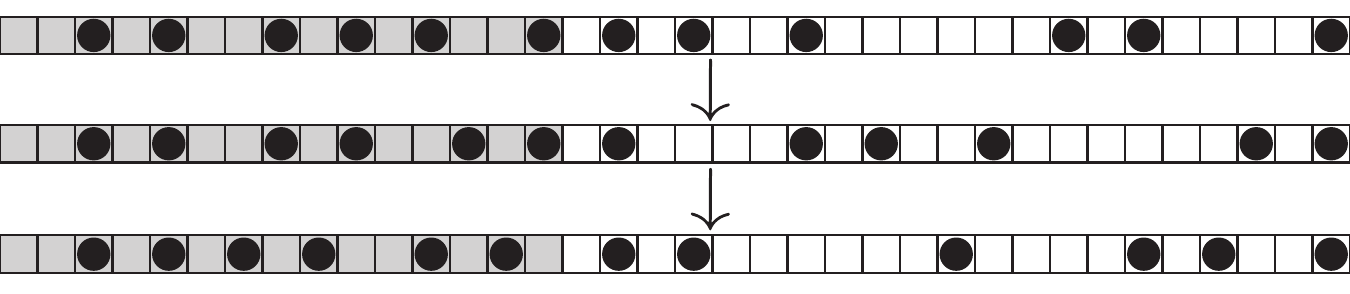}
\end{tabular}
\caption{We take snapshots for star marked states and consider the relation between snapshots.} \label{snapshotpart}
\end{figure}

\begin{figure}
\includegraphics[scale=0.70]{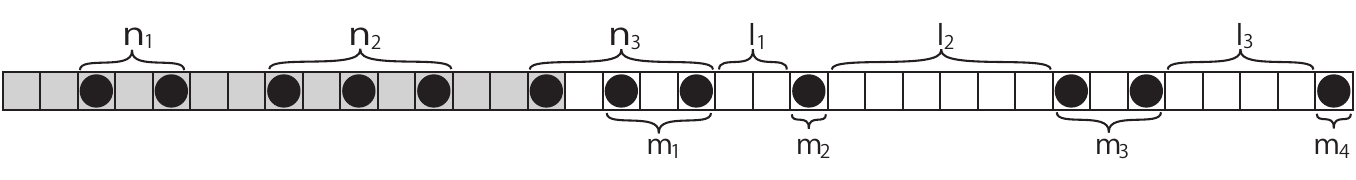}
\caption{The definition of $m$, $n$. We count $n$ and $m$ for the number of balls and $l$ the number of empty boxes. In this case, the data are $\alpha = 3$, $\gamma = 4$ and $(n_1, n_2, n_3; m_1, m_2, m_3, m_4; l_1, l_2, l_3) = (2, 3, 3; 2, 1, 2, 1; 2, 6, 4)$.} \label{fig3}
\end{figure}

To analyze such orbits, we introduce snapshots of the orbit when a notch arrives at the top of the main region (See Fig. \ref{snapshotpart}). From these snapshots, we can set $\alpha$ as the number of notches in the main region and $\gamma$ as the number of clusters in the skipped region. We also let $n_i$ be the number of balls between $i$-th and $i+1$-th notches for $i=1, \ldots, \alpha-1$ and $n_{\alpha}$ be that between $\alpha$-th notch and the first interval in the skipped region, $m_i$ be the number of balls of the $i$-th cluster (which is the same in the last section) and $l_i$ be the length of the intervals between $i$-th cluster in the skipped region and $i+1$-th one (See Figure \ref{fig3}). We note that balls in the $n_{\alpha}$ and $m_1$ are double-counted because it is convenient. After time evolutions, we can take the next snapshot and define $\{ n'_j \}$, $\{ m'_i \}$ and $\{ l'_i \}$ by the similar way.

These variables satisfy the relation:
\begin{eqnarray}
	\sum_{i=1}^\alpha n_i - m_1 = \frac{N_e - \alpha}{2} \label{reln1} \\ 
	\sum_{i=1}^\gamma m_i = M - \frac{N_e - \alpha}{2}
\end{eqnarray}
and the parities of $\alpha$ and $N_e$ are the same.

We next consider the relationship between $(n, m, l)$ and $(n', m', l')$. Now, we assume that the number of the notches and that of the clusters are preserved, i.e., there are no collision of clusters or no resolution of notches (we discuss the conditions later).

Since the clusters which are not last of each region are just traveling and make no collision, they just shift the order in the next snapshot. Then, we only focus on the last clusters. By the same discussion in the previous section, the last cluster in the main region separates balls staying around the main region and a new cluster in the skipped region and staying balls and the last cluster in the skipped region join together. The joining cluster of the skipped region push out the same amount of balls to the skipped region.
The intervals between the currently last cluster at the skipped region and the newly generated cluster is determined by the number of going around balls of the top cluster in the main region and the next notch. Therefore, the relationship between $(n, m, l)$ and $(n', m', l')$ expressed as:
\begin{eqnarray}
	n'_1 = n_\alpha - m_1 + m_\gamma \\
	n'_i = n_{i-1} \label{reln} \\
	m'_1 = m_\gamma \\
	m'_i = m_{i-1} \\
	l'_1 = 2(n_\alpha - m_1) + 2 \label{rellength} \\
	l'_i = l_{i-1}
\end{eqnarray}
We define this update map from $(n; m)$ to $(n'; m')$ as $F$, i.e., $(n'; m') = F(n; m)$, which is a linear one and its matrix representation is written in
\begin{equation}
	\left(\begin{array}{c} n'_1 \\ n'_2 \\ \vdots \\ n'_{\alpha-1} \\ n'_\alpha \\ \hline m'_1 \\ m'_2 \\ \vdots \\ m'_{\gamma-1} \\ m'_\gamma \end{array}\right)
	= \left(\begin{array}{ccccc|ccccc}   &  &  & & 1 & -1 &  & & & 1 \\
					      1 &  &  & & &  &  &  & & \\
					         & \ddots & & & & & & & & \\
					         & & 1 & & & & & & & \\
					         &  & & 1 & & &  & & & \\
				       \hline  &  & & &  &  &  &  & & 1 \\
				                  & & & &  &  1 &  & & & \\
				                  & & & & & & \ddots & & & \\
				                  & & & & & & & 1 & & \\
				                  & & & & & &  & & 1 & \end{array}\right)
				       \left(\begin{array}{c} n_1 \\ n_2 \\ \vdots \\ n_{\alpha-1} \\ n_\alpha \\ \hline m_1 \\ m_2 \\ \vdots \\ m_{\gamma-1} \\ m_\gamma \end{array}\right).
\end{equation}
The eigenvalues of this matrix are $e^{2\pi i \mu/\alpha}$ ($\mu=0, 1, \ldots, \alpha$) and $e^{2\pi i \nu/\gamma}$ ($\nu=0, 1, \ldots, \gamma$). 

If $\alpha$ and $\gamma$ are co-prime, this matrix has eigenvalue $1$ with algebraic multiplicity $2$ but the dimension of corresponding eigenspace is also $2$, that is, arbitrary initial states is decomposed by linear combination of the eigenvector. Then, by applying this map $\alpha \gamma$ times, the data finally become the same as the initial data, which is consistent with the periodicity of the system.

However, if $\alpha$ and $\gamma$ are not co-prime, $e^{2\pi i v/q}$ ($v=1,2,\ldots, q-1$) are also eigenvalues with multiplicity $2$ but the dimension of the eigenspace is 1, where $q = \gcd(\alpha, \gamma)$. Then, there can be an initial state which is not expressed as a linear combination of the eigenvectors. If we take such an initial state, the system seems to behave not periodic and the data $(n; m)$ seem to grow as polynomial order with oscillation. However, the conditions to preserve the form of cluster break after several steps. Then, we do not consider such cases. The condition which an initial state is expressed as a linear combination of the eigenvectors is written in
\begin{eqnarray}
	m_1 + m_{1+q} + \ldots + m_{1+(d-1)q} = m_2 + m_{2+q} + \ldots + m_{2+(d-1)q} = \ldots = m_q + m_{2q} + \ldots + m_{dq} \label{condmulti}
\end{eqnarray}
where $d = \gamma/q$.

\bigskip

We now consider the conditions to preserve the form of clusters which is assumed in this discussion. It is required that the all clusters are in the system. Then, the one has:
\begin{equation} \label{cond1}
	1 + \sum_{i=1}^\gamma (2m_i - 1) + \sum_{i=1}^{\gamma-1} l_i \le N_i.
\end{equation}

Similarly to the discussion in the previous section, we next consider characteristic time which determines the behavior of the system:
\begin{enumerate}
\item The top cluster at the skipped region arrives at the last site of the skipped region:
\begin{equation}
	t_0 = N_s - \left(1+\sum_{i=1}^{\gamma} (2 m_i - 1) + \sum_{i=1}^{\gamma-1} l_i\right).
\end{equation}
\item The top notch arrives at the last site of the main region:
\begin{equation}
	t_1 = 2(n_\alpha - m_1) - 1.
\end{equation}
\item All the balls of the top cluster go back to the main region:
\begin{equation}
	t_2 = t_1 + 2 m_\gamma - 1 = 2 n_\alpha - 2.
\end{equation}
\item The second top cluster arrives at the last site of the skipped region:
\begin{equation}
	t_3 = N_s - \left(1+\sum_{i=1}^{\gamma-1} (2 m_i - 1) + \sum_{i=1}^{\gamma-2} l_i\right).
\end{equation}
\item The second notch arrives at the last of the main region:
\begin{equation}
	t_4 = 2(n_\alpha - m_1 + n_{\alpha-1}).
\end{equation}
\end{enumerate}

To preserve the number of the notches and the clusters, this characteristic time must satisfy the relation $(0 \le ) t_0 \le t_1<t_2<t_3\le t_4$. If not, the shape of the orbit is finally broken. For example, if $t_1 < t_0$, we have to add the cluster with length 0 like that in subsection E, if $t_4 < t_3$, notches finally resolve and if $t_5 < t_3$, the second top cluster catches up the top one before it goes to main region and two clusters coalesce.

However, these conditions guarantee only the existence of the next data and we need to check that the new data also satisfy those conditions again to construct periodic orbits. Repeating this procedure at most $\alpha \gamma$ times, we can check that the given initial data generate the periodic orbit. 

\bigskip

To calculate the period and the flow, we consider the value$\sum_{s=0}^{\alpha \gamma - 1}n^{(s)}_1$, where $(n^{(0)}; m^{(0)})$ is the initial data $(n; m)$ and $(n^{(s)}, m^{(s)}) = F(n^{(s-1)}, m^{(s-1)})$ recursively. By virtue of the rule (\ref{reln}) and (\ref{reln1}):
\begin{eqnarray}
	\sum_{s=0}^{\alpha-1} n^{(s)}_1 - m_{\gamma-\alpha+2} =  \sum_{s=0}^{\alpha-1} n^{(\alpha-1)}_{1+s}  - m_{\gamma-\alpha+2} = \frac{N_e-\alpha}{2},
\end{eqnarray}
we obtain
\begin{eqnarray}
	\sum_{s=0}^{\alpha \gamma -1} n^{(s)}_1 = \sum_{r=0}^{\gamma-1} \left(\sum_{s=0}^{\alpha - 1} n^{(r \alpha + s)}_1 \right) = \sum_{r=0}^{\gamma-1} \left(\sum_{s=0}^{\alpha - 1} n^{(r \alpha)}_{1+s} \right) = \sum_{r=0}^{\gamma-1} \left(\frac{N_e-\alpha}{2} + m_{\gamma - r \alpha + 2} \right),
\end{eqnarray}
where the index of $m$ takes modulo of $\gamma$. If $\gamma$ and $\alpha$ are co-prime, $\gamma - r \alpha + 2$ ($r=0, 1, \ldots, \gamma - 1$) runs on all of $1, 2, \ldots, \gamma$ just at once. If not, $\gamma - r \alpha + 2$ ($r=0, 1, \ldots, \gamma -1$) runs on $\gamma - q + 2, \gamma - 2q + 2, \ldots, \gamma - dq + 2$ at $q$ times, where $\gcd(\alpha, \gamma) = q$ and $d = \gamma /q$. However, by virtue of the condition (\ref{condmulti}), finally for both cases, one has
\begin{equation}
	\sum_{r=0}^{\gamma-1} m_{\gamma - r \alpha + 2} = \sum_{i=1}^{\gamma} m_i = M - \frac{N_e - \alpha}{2} \label{msum}
\end{equation}
and
\begin{eqnarray}
	\sum_{s=0}^{\alpha \gamma - 1} n^{(s)}_1 = \gamma \frac{N_e - \alpha}{2} + M - \frac{N_e - \alpha}{2} = \frac{2 M + (\gamma - 1)( N_e - \alpha)}{2}. \label{sumnformula}
\end{eqnarray}

The period of the oribit is calculated by summing the time to transit each snapshot. By virtue of the similar discussion in the section C, the time to transit from the initial state to the next state satisfying $u^t_0 = u^t_1 = 0$ is $2 n_1 + 1$.  Then, if $\gcd(\alpha, \gamma) = 1$, the period is 
\begin{equation}
	T = \sum_{s=0}^{\alpha \gamma - 1} (2 n^{(s)}_1 + 1).
\end{equation}
and due to (\ref{sumnformula}), the period of this orbit is expressed as
\begin{equation}
	T = 2 M + (\gamma - 1) N_e + \alpha
\end{equation}
In the case of $\gcd(\alpha, \gamma) = q > 1$, the real period of this system obtained by $T/q$ because the same snapshot update repeats just $q$ times.

The flow is also calculated by summing the number of emitted balls from the main region to the skipped region for transitioning each snapshot and dividing this sum by the period and finally written in
\begin{eqnarray} 
	\langle J_s \rangle = \frac{\sum_{s=0}^{\alpha \gamma - 1} m^{(s)}_1}{T} =  \frac{\alpha(2M - N_e + \alpha)}{2(2M + (\gamma - 1) N_e + \alpha)} \label{flow-i-a-g}  \\
	\langle J_m \rangle = \frac{\sum_{s=0}^{\alpha \gamma - 1} n^{(s)}_1}{T} = \frac{2M + (\gamma - 1) (N_e - \alpha)}{2(2M + (\gamma - 1) N_e + \alpha)},
\end{eqnarray}
which agrees the result of the numerical simulation. These formulae of the flow preserves even if the real period is shorter.

The formulae of the flow depends only on the number of balls, clusters and notches and the length of the main region, similarly to those in the last section. However, the condition depends on the detail of the data.

\begin{figure}
\includegraphics[]{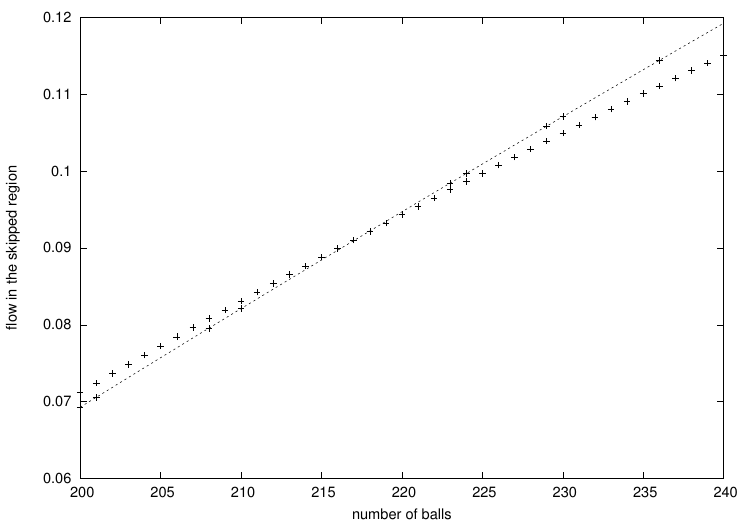}
\caption{A part of the graph of the flow in the skipped region for $N_m = 301$, $N_s = 700$. The dashed curve are by (\ref{flow-i-a-g}) for $\alpha = 3$, $\gamma = 7$.} \label{fig-expr-vs-theor-special}
\end{figure}

FIG \ref{fig-expr-vs-theor-special} is partially magnified of FIG \ref{fig_flow_skipped}. The dashed line is the theoretical value of the flow of general periodic orbits and several points are out of this curve, which corresponds the special periodic orbits. The dotted line is that of the special periodic orbits for $\alpha = 3$ and $\gamma = 7$. Therefore, we can confirm that those exceptional data correspond the special periodic orbits with 3 notches and 7 clusters.

The result in this section are finally reduced to that in the previous section when setting $\alpha=1$.

\begin{figure}
\includegraphics[scale=0.75]{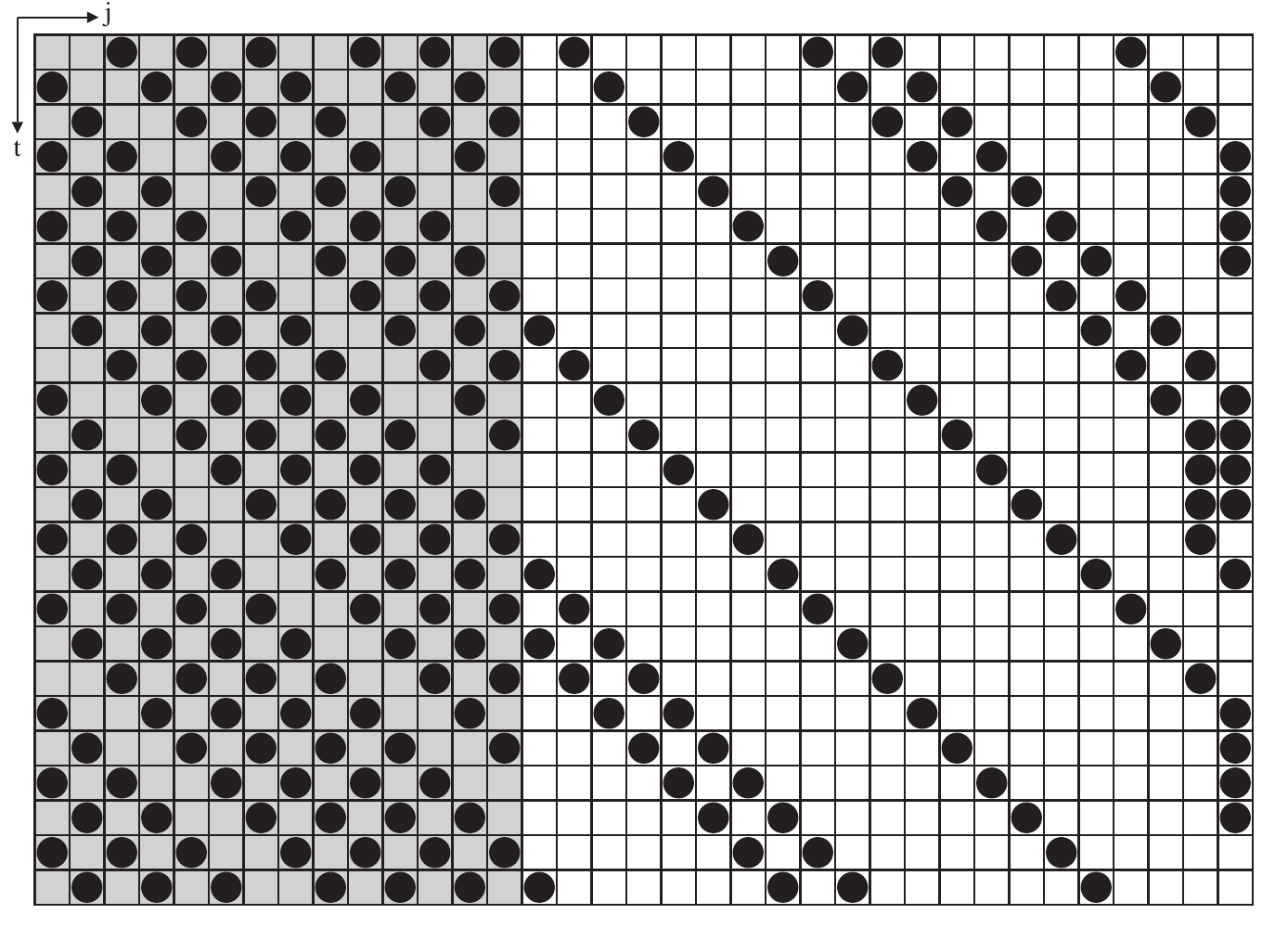}
\caption{An example of special limit cycles for the even length main region. The system parameters are $N_e = 14$, $N_i = 21$, $M = 10$.} \label{fig-smpl-time-evol4}
\end{figure}

We can also apply this approach when $N_m$ is even. In this case, the balls in the skipped region can go back to the main region differently from the general orbit (See Fig. \ref{fig-smpl-time-evol4}). The limit cycle where the balls at the skipped region creates jams and never go back to main region is considered as that in the case $\alpha=0$.

The evolution rule of data $n$ and $m$ also holds even if some of $n$ and $m$ take $0$. However, the correspondence between the state $u^t_j$ and the data becomes more complicated for such cases. For example, the initial state in the Figure \ref{examplespecial} seems to have 2 notches and 1 cluster but $\alpha = 3$ and $\gamma = 5$ (under this settings, the evolution of the system is well explained). The relation (\ref{rellength}) helps to obtain the real number of the notches and clusters. If the real length of interval between two clusters is different from the length obtained from (\ref{rellength}), there is the position where null clusters should be inserted. 

\section{Dynamics with several skipped regions}

\begin{figure}
\begin{center}
\includegraphics[scale=0.5, angle=90]{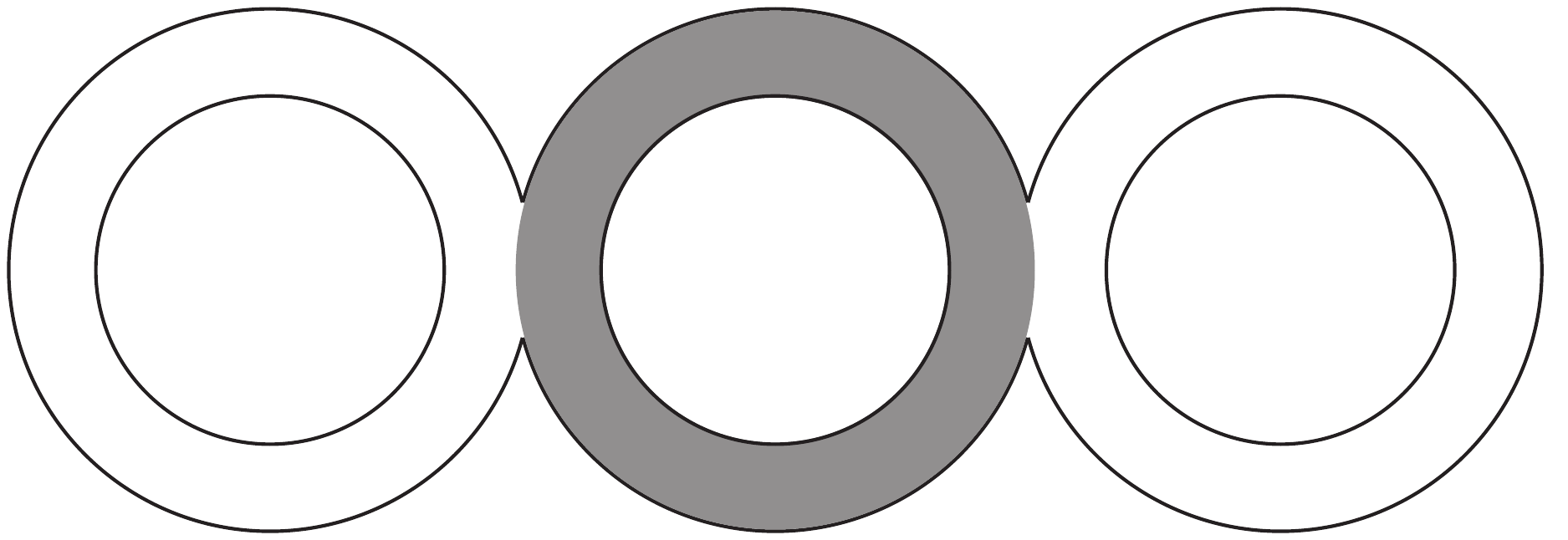}
\caption{A topological representation of two main and skipped region system.} \label{3rings}
\end{center}
\end{figure}

Our approach in the previous section is also applicable to analyze periodic orbits for the system for $K \ge 2$ (i.e. with several jumping sites). Observing the dynamics of orbits shows that we set the sites from $\iota_{k-1}$-th to $\epsilon_k$-th one as $k$-th main region and from $(\epsilon_k+1)$-th to  $(\iota_k-1)$-th as $k$-th skipped region for $1 \le k \le K$ which is similar to the case of $K=1$. This means that there are several main and skipped regions. We note that the number of main regions and skipped regions are trivially the same.
In fact, the topology of the path for this system is depicted in FIG. \ref{3rings} and the system consists on one main region and two detours in mid-course of the main region.

\begin{figure}
\begin{center}
\includegraphics[scale=0.6]{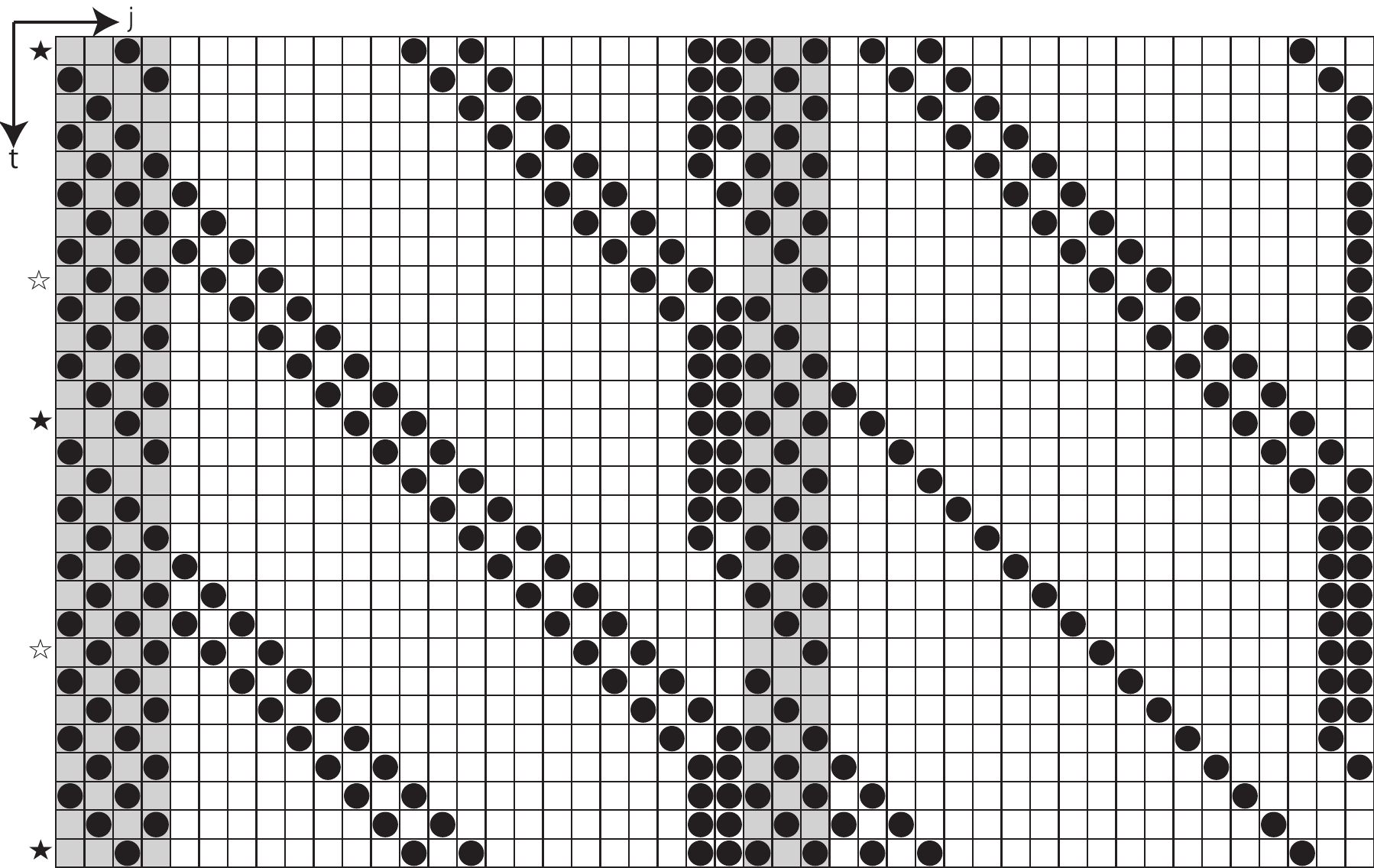}
\caption{An example of periodic orbits for two main and skipped regions. The time when the notch arrived at the top of the first main region is filled-star marked and that when the notch arrived at the top of the second is outline-star marked.} \label{tworegions3}
\end{center}
\end{figure}

\begin{figure}
\begin{center}
\includegraphics[scale=0.6]{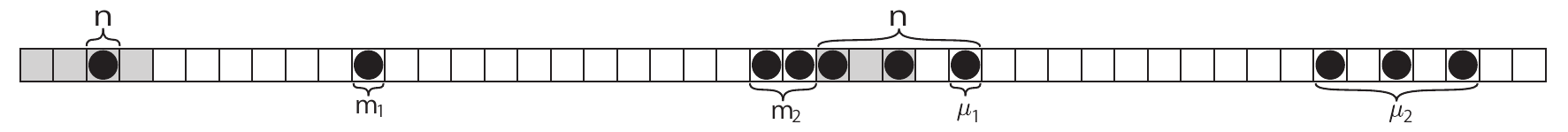}
\caption{Definition of the data for the filled-star marked states. In this example, the data is $(n; m_1, m_2; \mu_1, \mu_2) = (4; 1, 2; 1, 3)$.} \label{snapshot1}
\end{center}
\end{figure}

\begin{figure}
\begin{center}
\includegraphics[scale=0.6]{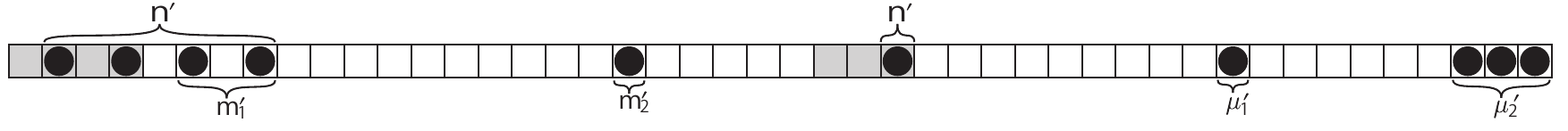}
\caption{Definition of the data for the outline-star marked states. In this example, the data is $(n'; m'_1, m'_2; \mu'_1, \mu'_2) = (5; 2, 1; 1, 3)$.} \label{snapshot2}
\end{center}
\end{figure}

Let us construct the general periodic orbits. 
Similarly to the discussion in the last section, we focus on the behavior of the notch and take snapshots of orbits when it arrives at the top of each main region and we consider the mappings for transition between these snapshots. For example, in the case of $K=2$,  we take two types of snapshots of an orbit (See figure \ref{tworegions3}). We define $n$ as the number of balls forming the cluster in the main region and $m_i$ as that forming $i$-th cluster in the first skipped region and $\mu_i$ as that forming $i$-th cluster in the second skipped region. (See Figures \ref{snapshot1} and \ref{snapshot2})

If the initial data $(n; m_1, \ldots, m_{\gamma}; \mu_1, \ldots, \mu_{\gamma'})$ is given, the data $(n'; m'_1, \ldots, m'_\gamma; \mu'_1, \ldots, \mu'_{\gamma'})$ is expressed as
\begin{equation}
	\left(\begin{array}{c} n' \\ \hline m'_1 \\ m'_2 \\ \vdots \\ m'_{\gamma-1} \\ m'_\gamma \\ \hline \mu'_1 \\ \mu'_2 \\ \vdots \\ \mu'_{\gamma'-1} \\ \mu'_{\gamma'} \end{array}\right)
	= \left(\begin{array}{c|ccccc|ccccc} 1 & & & & & 1 & -1 & & & & \\
					\hline  & 1 &  &  & & &  &  & & & \\
					       & & 1 &  & & &  &  &  & & \\
					        & & & \ddots & & & & & & & \\
					        & & & & 1 & & & & & & \\
					        & &  & & & 1 & &  & & & \\
				       \hline  & &  & & &  & &  &  & & 1 \\
				                 & & & & &  & 1 & & & & \\
				                 & & & & & & & \ddots & & & \\
				                 & & & & & & & & 1 & & \\
				                 & & & & & & &  & & 1 & \end{array}\right)
				       \left(\begin{array}{c} n \\ \hline m_1 \\ m_2 \\ \vdots \\ m_{\gamma-1} \\ m_\gamma \\ \hline \mu_1 \\ \mu_2 \\ \vdots \\ \mu_{\gamma'-1} \\ \mu_{\gamma'} \end{array}\right).
\end{equation}
Let the data when the notch which departed from the second main region arrives at the first main region be $(\tilde{n}; \tilde{m}_1, \ldots, \tilde{m}_\gamma; \tilde{\mu}_1, \ldots, \tilde{\mu}_{\gamma'})$ and the relation between these data and  $(n'; m'_1, \ldots, m'_\gamma; \mu'_1, \ldots, \mu'_{\gamma'})$ is written in
\begin{equation}
	\left(\begin{array}{c} \tilde{n} \\ \hline \tilde{m}_1 \\ \tilde{m}_2 \\ \vdots \\ \tilde{m}_{\gamma-1} \\ \tilde{m}_\gamma \\ \hline \tilde{\mu}_1 \\ \tilde{\mu}_2 \\ \vdots \\ \tilde{\mu}_{\gamma'-1} \\ \tilde{\mu}_{\gamma'} \end{array}\right)
	= \left(\begin{array}{c|ccccc|ccccc} 1 & -1 & & & & & & & & & 1 \\
					\hline  &  &  &  & & 1 &  &  & & & \\
					       & 1 &  &  & & &  &  &  & & \\
					        & & \ddots & & & & & & & & \\
					        & & & 1 & & & & & & & \\
					        & &  & & 1 & & &  & & & \\
				       \hline  & &  & & &  & 1 &  &  & & \\
				                 & & & & &  &   & 1 & & & \\
				                 & & & & & & &  & \ddots & & \\
				                 & & & & & & & &  & 1 & \\
				                 & & & & & & &  & &  & 1 \end{array}\right)
				       \left(\begin{array}{c} n' \\ \hline m'_1 \\ m'_2 \\ \vdots \\ m'_{\gamma-1} \\ m'_\gamma \\ \hline \mu'_1 \\ \mu'_2 \\ \vdots \\ \mu'_{\gamma'-1} \\ \mu'_{\gamma'} \end{array}\right).
\end{equation}
Then, when the notch comes back to the top of the first main region again, the matrix representation of the update of $m$ and $n$ is written in the product of two matrices representations expressed as 
\begin{equation}
\left(\begin{array}{c} \tilde{n} \\ \hline \tilde{m}_1 \\ \tilde{m}_2 \\ \vdots \\ \tilde{m}_{\gamma-1} \\ \tilde{m}_\gamma \\ \hline \tilde{\mu}_1 \\ \tilde{\mu}_2 \\ \vdots \\ \tilde{\mu}_{\gamma'-1} \\ \tilde{\mu}_{\gamma'} \end{array}\right)
	= \left(\begin{array}{c|ccccc|ccccc} 1 & -1 & & & & 1 & & & & & \\
					\hline  &  &  &  & & 1 &  &  & & & \\
					       & 1 &  &  & & &  &  &  & & \\
					        & & \ddots & & & & & & & & \\
					        & & & 1 & & & & & & & \\
					        & &  & & 1 & & &  & & & \\
				       \hline  & &  & & &  & & &  & & 1 \\
				                 & & & & &  & 1 & & & & \\
				                 & & & & & & & \ddots & & & \\
				                 & & & & & & & & 1 & & \\
				                 & & & & & & &  & & 1 & \end{array}\right)
				                 \left(\begin{array}{c} n \\ \hline m_1 \\ m_2 \\ \vdots \\ m_{\gamma-1} \\ m_\gamma \\ \hline \mu_1 \\ \mu_2 \\ \vdots \\ \mu_{\gamma'-1} \\ \mu_{\gamma'} \end{array}\right)
\end{equation}
The eigenvalues of this matrix are all roots of unity and the dimension of each eigenspace is equal to its algebraic multiplicity.

\begin{figure}
\begin{center}
\includegraphics[scale=0.6]{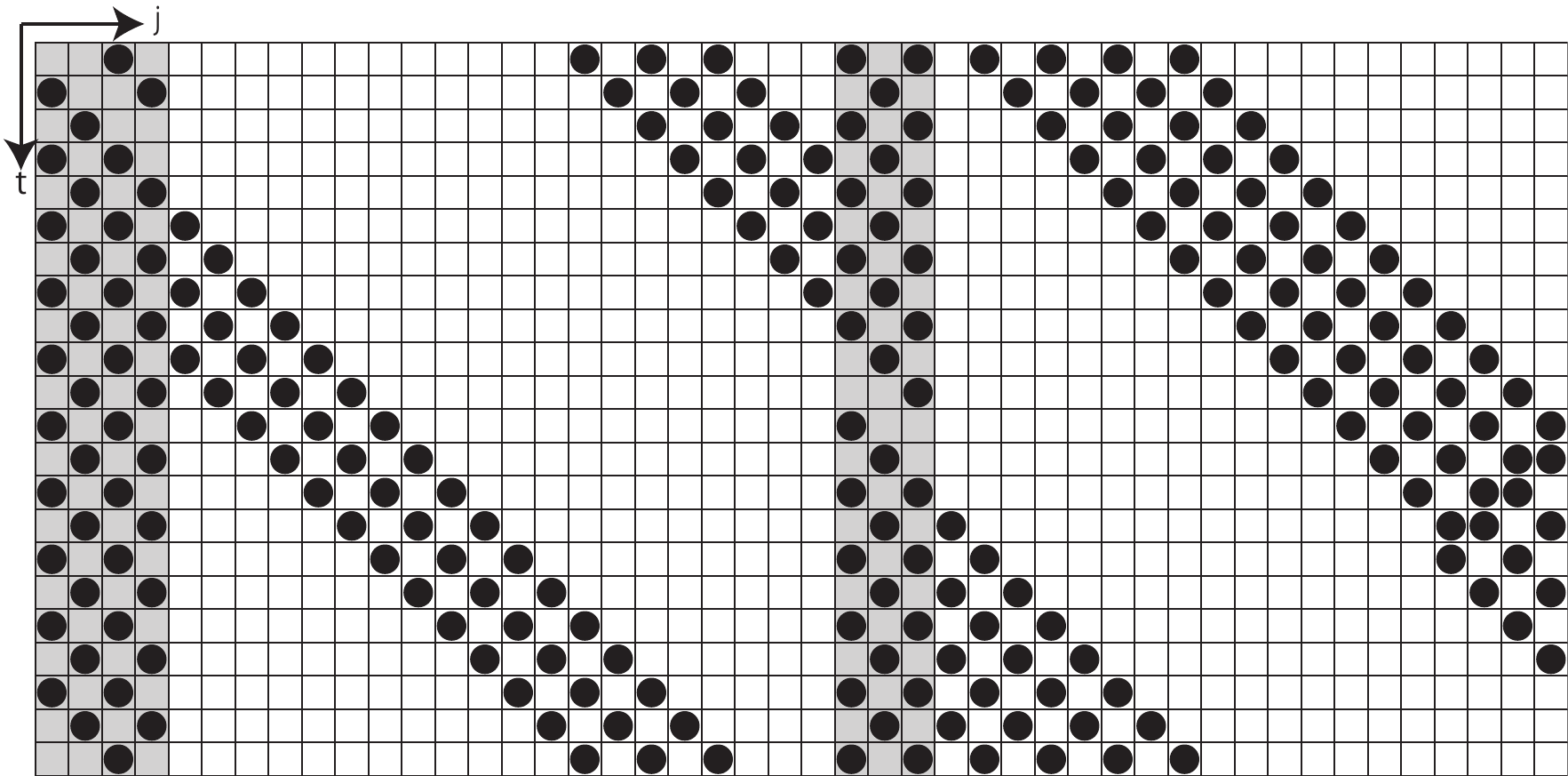}
\caption{Another example of periodic orbits for two main and skipped regions. The system parameters and the number of balls are the same as those of Fig. \ref{tworegions3} and only one notch appears in the main region. However, the number of clusters are different.} \label{tworegions1}
\end{center}
\end{figure}

By virtue of the similar approach in the previous sections, we can also obtain the flow and period of these periodic orbits and the condition that orbits hold. However, differently from the case $K=1$, the flow depends on the length of the skipped regions and the detail of the data. Furthermore, we observed that there is a possibility that two general periodic orbits co-exists (See Fig. \ref{tworegions1}). 

This approach is difficult to be applied for special periodic orbits for which there appear several clusters in the main region because the notch affecting the dynamics depends on the initial state and it is difficult to determine such a notch. We believe that this difficulty is the equivalent as to analyze the orbit directly.

\section{Concluding Remarks}

In this paper, we considered the dynamics for the simplified version of the path-preference model. We discuss the property for a class of limit cycles, which are raised by general initial states, for example, the conditions of system parameters where the orbit is attracted by one of them and the exact value of the flow in each part which depend only on the length of each part and the number of balls. 

We pay attention the state where the notch in the main region is at the top. That is, we focus on the subset of states $S_0 = \{ u \in \{0, 1\}^N \mid u_0 = u_1 = 0 \}$ and express the change of the state between two states which an orbit started from the intersection with $S_0$ and the next intersection after this orbit departed from $S_0$ to understand the behavior of the orbit. This is nothing but the cellular automaton analogue of the idea of Poincar\'e section and Poincar\'e map in the continuous time dynamical systems.

Generally, it is difficult to apply the concept of Poincar\'e section to discrete dynamical systems naively because the orbit is discrete. However, in cellular automata, dependent variables are also discrete. Therefore, the analogue of the section can be defined naturally because there is a case where a discrete analogue of continuity of orbits can be defined naturally.

We set the time evolution rule of the system deterministic and containing a symmetry. Due to such settings, we have to mention that the behavior of orbits becomes different by the parity for the length of the main region. Therefore, it is hard to say that our analysis result can apply directly to that of the real dynamics for living systems.
However, by customizing the settings of this system, we can consider other cellular automaton models as the variations to weaken such a parity and it is expected that the dynamics of these models can apply to analysis for real dynamics.

For example, let us introduce the move prior rule which is described in Section 2. The flow does not seem to change discontinuously by the number of balls in the numerical simulation. However, the dynamics is more complex than that of the jump prior rule. There is a limit cycle with ten billion order period for $N_m=30$ and $N_s=71$. To take such a long period, the dynamics contains some complicated structures.

\begin{figure}
\includegraphics[scale=0.8]{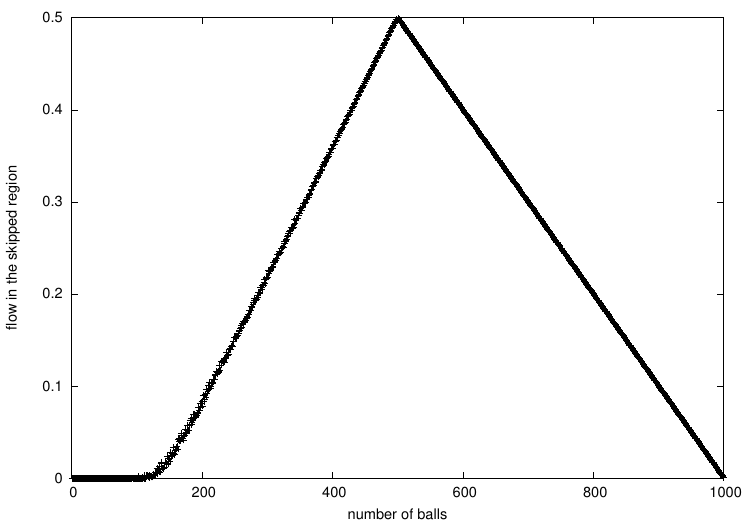}
\caption{The graph of the flow in the skipped region $\langle J_s \rangle$ for the move prior rule path-preference cellular automaton where $N_m = 301$ and $N_s = 700$.} \label{fig_flow_moveprior}
\end{figure}

We can also introduce some stochastic factors for this model. The most famous one is in the moving process to the next site. However, we consider the case where the ball at the end of the main region jumps to the top of it or moves to the top of the skipped region stochastically. In this case, the flow loses discontinuousness for the number of balls. If jump probablity $p=0$, the dynamics is nothing but that of the TASEP and if $0<p<1$, the dynamics behaves intermediate one between the TASEP ($p=0$) and that described mainly in this paper ($p=1$). However, the analysis for this model is further difficult problem because we cannot introduce the mean field approximation method like that of the ASEP and Statistical dynamics methods require sufficient length of the system to ignore the effect of special sites, for example, bifurcating or confluence ones.

\section*{Acknowledgements}
We would like to thank Professor Tetsuji Tokihiro and Professor Youichiro Wada for helpful comments. This research is supported by Platform for Dynamic Approaches to Living System (the Platform Project for Supporting in Drug Discovery and Life Science Research) from the Ministry of Education, Culture, Sports, Science (MEXT) and Technology, Japan, and Japan Agency for Medical Research and Development (AMED). This work is also partially supported by the JPSJ KAKENHI Grant Number 17K14199.

\bibliographystyle{unsrt}

\end{document}